\documentclass[aps,showpacs,amsmath,amssymb,twocolumn,prl,superscriptaddress,notitlepage]{revtex4-2}

\usepackage{qcircuit}
\usepackage{amsmath,bm}
\usepackage[dvips]{graphicx}
\usepackage{amsmath,amssymb,amsthm,mathrsfs,amsfonts,dsfont}
\usepackage{subfigure, epsfig}
\usepackage{braket}
\usepackage{bm}
\usepackage{enumerate}
\usepackage{color}
\usepackage{graphicx}
\usepackage{algorithm}
\usepackage{algorithmic}
\usepackage{braket}
\usepackage{comment}
\usepackage{appendix}
\usepackage{here}
\usepackage{tabularx}
\usepackage{physics}
\usepackage{mathtools}

\definecolor{grey}{rgb}{.5,.5,.5}

\newcommand{\ts}[1]{#1}

\begin{document}


\title{Non-Markovianity in Quantum Information Processing: \\
Interplay with Quantum Error Mitigation }



\author{Suguru Endo}
\email{suguru.endou@ntt.com}
\thanks{These authors contributed equally to this work.}
\affiliation{NTT Computer and Data Science Laboratories, NTT Inc. 3-9-11 Midori-Cho Musashino 180-8585, Japan}
\affiliation{NTT Research Center for Theoretical Quantum Information, NTT Inc. 3-1 Morinosato Wakanomiya, Atsugi, Kanagawa, 243-0198, Japan}

\author{Hideaki Hakoshima}
\email{hakoshima.hideaki.es@osaka-u.ac.jp}
\thanks{These authors contributed equally to this work.}
\affiliation{Graduate School of Engineering Science, The University of Osaka, 1-3 Machikaneyama, Toyonaka, Osaka 560-8531, Japan}
\affiliation{Center for Quantum Information and Quantum Biology, The University of Osaka, 1-2 Machikaneyama, Toyonaka, Osaka 560-0043, Japan}

\author{Tomohiro Shitara}
\email{tomohiro.shitara@ntt.com}
\affiliation{NTT Computer and Data Science Laboratories, NTT Inc. 3-9-11 Midori-Cho Musashino 180-8585, Japan}
\affiliation{NTT Research Center for Theoretical Quantum Information, NTT Inc. 3-1 Morinosato Wakanomiya, Atsugi, Kanagawa, 243-0198, Japan}


\begin{abstract}
Non-Markovian dynamics are typically present in the dynamics of open quantum systems. Despite the rich structure of non-Markovian dynamics, their relevance to quantum information processing (QIP) has been rarely discussed. In this work, we demonstrate that a characteristic of non-Markovian dynamics naturally arises in quantum error correction (QEC) and quantum teleportation. The non-Markovianity in open quantum systems is naturally attributed to the information backflow from the environment. We partition the whole Hilbert space into the logical subsystem and the gauge subsystem. The logical subsystem stores the quantum information for QIP, while the gauge subsystem stores the information for recovery of the logical information, i.e., the syndrome measurement outcomes for QEC and Bell measurement outcomes for successful teleportation. We then show that the non-Markovianity in QIP appears as a consequence of the feedback operation based on the measurement outcomes of the gauge subsystem. Finally, we show that the non-Markovianity in QIP reduces the sampling cost of quantum error mitigation (QEM), shedding light on the importance of combination strategies of QEC and QEM in a practical QIP.

\end{abstract}

\maketitle
\emph{Introduction.---} 
The formulation of open quantum systems plays a fundamental role in quantum information processing (QIP) because the system of interest inevitably interacts with the environment, which generally results in computation errors~\cite{nielsen2010quantum,gardiner2004quantum}. The noise effect is often modeled by 
\ts{Gorini–Kossakowski–Sudarshan–Lindblad (GKSL)} master equations, in many theoretical analyses~\cite{nielsen2010quantum,gardiner2004quantum}. However, the memory effect between the system and environment generally exists, necessitating the description of the non-Markovian system dynamics~\cite{breuer2016colloquium,rivas2014quantum}. 

In the presence of the memory effect, quantum information can flow back to the system from the environment~\cite{breuer2016colloquium,rivas2014quantum}. Previous works \cite{rivas2010entanglement,hall2014canonical} have shown that the non-Markovian dynamics can be captured by the canonical master equation form:
$\frac{d}{dt}\rho(t) = \sum_k \gamma_k (t) [L_k(t) \rho(t) L_k(t)^\dag - \frac{1}{2} \{L_k^\dag (t) L_k (t), \rho(t) \} ]$, where we neglect the Hamiltonian term for simplicity
~\cite{hall2014canonical}. 
This equation has the same form as the \ts{GKSL} master equation, but the decay rates $\gamma_k (t)$ can take negative time-dependent values, which indicates the presence of non-Markovianity.
While many previous studies have considered non-Markovian effects in the physical situations that realize QIP, including quantum error correction (QEC), quantum error mitigation (QEM), and feedback controls under non-Markovian noise \cite{mann2025quantum,puviani2025non,Biswas_2025,cao2023quantum,liu2024non,swain2025noise}, almost no research has focused on revealing non-Markovian effects intrinsic in QIP itself.

\begin{figure*}[htbp]
    \centering
    \includegraphics[width=\linewidth]{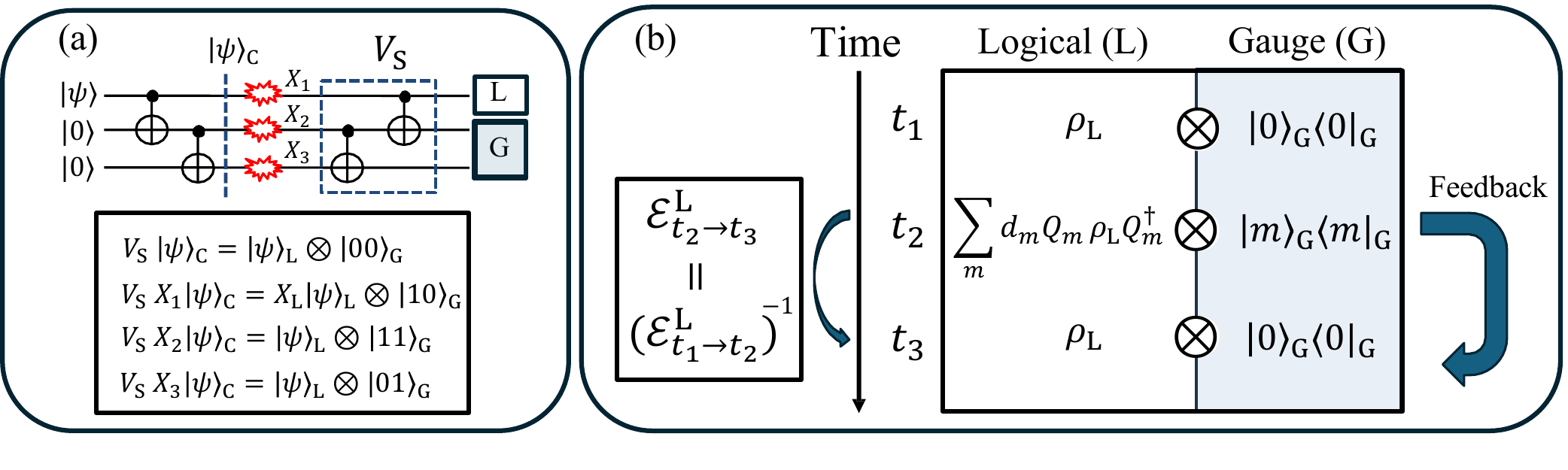}
    \caption{The conceptual figure illustrating an example of our theory. (a) The subsystem unitary isomorphism operator $V_{\rm S}$ to decompose the state into the logical and gauge subsystems, illustrated by the three-qubit code. In this case, $V_{\rm S}$ is a Clifford unitary operator. (b) The emergence of the non-Markovian processes in QEC. First, the logical state is initialized in $\rho_{\rm L}$ at time $t_1$. The reduced density matrix in the logical subsystem changes into a mixed state $\sum_m d_m Q_m \rho_{\rm L} Q_m^\dag$ at time $t_2$ but correlates with the gauge subsystem. We can perform a feedback operation based on the gauge information, and the state changes into the original state $\rho_{\rm L}$ at time $t_3$ when the KL condition is satisfied. Then, the quantum process from $t_2$ to $t_3$ is generally a non-CPTP map, indicating the presence of non-Markovianity.
    }
    \label{Fig: concept}
\end{figure*}

In this work, we show that non-Markovianity is an inherent and necessary feature of the effective dynamics during QIP protocols, arising from their structure. Unlike the harmful noise effect of external environments, this intrinsic non-Markovianity describes the mechanism of information recovery within the QIP protocol. We introduce the following examples: Pauli-based QEC~\cite{gottesman2009introduction}, bosonic quantum error correction~\cite{cai2021bosonic,ma2021quantum}, and quantum teleportation~\cite{bennett1993teleporting,pirandola2015advances}. We present a conceptual figure illustrating non-Markovianity in our theoretical framework in Fig. \ref{Fig: concept}, referring to the example of the three-qubit code. The key point is that we introduce the subsystem frame and divide the total system into logical subsystems that have quantum information for \ts{QIP}, and the gauge subsystem with the information of the syndrome or the Bell measurement. In this picture, the logical subsystem and the gauge subsystem are generally correlated, but they can be decoupled using the measurement and feedback operations. We show that the dynamics of the logical subsystem can be generally represented by the non-completely positive (non-CP) maps, with the continuous dynamics being described by the canonical form of the master equation with negative coefficients when tracing out the gauge subsystem.

We further explore the interplay between non-Markovian dynamics in QEC and QEM~\cite{temme2017error,endo2018practical,endo2021hybrid,cai2023quantum,hakoshima2021relationship}. QEM suppresses computation errors at the cost of additional repetitions of quantum computation, known as the QEM sampling cost. The QEM sampling cost reduction due to QEC has been demonstrated under the specific choice of QEM~\cite{suzuki2022quantum,xiong2020sampling}. Considering the fundamental limitation of QEM described by the distance measure~\cite{takagi2022fundamental,takagi2023universal}, we show that the fundamental QEM cost bound can be improved depending on the amount of a non-Markovian measure, i.e., the decay rate measures. While it has been shown that the non-Markovianity helps reduce the QEM costs in a specific QEM protocol~\cite{hakoshima2021relationship}, our results indicate that the non-Markovianity due to QEC reduces the fundamental QEM cost for mitigating errors in the logical subsystem.

\emph{Non-Markovianity.---}
We review the non-Markovian dynamics in quantum processes~\cite{breuer2016colloquium,rivas2014quantum}. Let the initial state be denoted as $\rho_{\mathrm{S} } \otimes \rho_{\mathrm{E} }$, where $\rm{S}$ and $\rm{E}$ correspond to the system and the environment. Let us denote the three time points $t_1$, $t_2$, and $t_3$ with $t_1 < t_2 < t_3$. Due to the unitary interaction $U^{(t_1, t_2)}_{\rm SE}$ between the system and the environment during the time interval $[t_1, t_2]$, the system state changes into $\mathcal{E}_{t_1, t_2}(\rho_{\rm S})=\mathrm{Tr}_{\rm E}[U^{(t_1, t_2)}_{\rm SE} \rho_{\rm S} \otimes \rho_{\rm E} U^{(t_1, t_2 ) \dag}_{\rm SE}]$, where $\mathcal{E}_{t_1, t_2}$ denotes the process the system state undergoes. Meanwhile, for the unitary interaction in the time interval $[t_1, t_3]$, we have $\mathcal{E}_{t_1, t_3}(\rho_{\rm S})=\mathrm{Tr}_{\rm E}[U^{(t_1, t_3)}_{\rm SE} \rho_{\rm S} \otimes \rho_{\rm E} U^{(t_1, t_3 ) \dag}_{\rm SE}]$. 
If $\mathcal{E}_{t_1,t_2}^{-1}$ is invertible and the dynamical map $\mathcal{E}_{t_2, t_3}= \mathcal{E}_{t_1,t_3}\circ (\mathcal{E}_{t_1,t_2})^{-1}$ is a completely positive map for all $t_2\in (t_1, t_3)$,
$\mathcal{E}_{t_1, t_3}$ is called \textit{CP-divisible} map, and widely accepted as a definition of Markovian processes~\cite{rivas2010entanglement}. Otherwise, the processes are non-Markovian. 

While many measures have been proposed as a metric of non-Markovianity, we mainly focus on \ts{the} Rivas-Huelga-Plenio (RHP) measure~\cite{rivas2010entanglement} defined as $\mathcal{N}_{\rm RHP}(t_1\rightarrow t_3)=\int_{t_1}^{t_3} n(t)dt$ for $n(t)= \mathrm{lim}_{\delta \rightarrow 0} (\| (\mathcal{E}_{t, t+\delta}\otimes \mathcal{I}) [\ket{\Psi}\bra{\Psi}] \|_1-1)/\delta $, where $\|\cdot\|_1$ denotes the trace norm and $\ket{\Psi}\propto \sum_i\ket{ii}$ is the maximally entangled state between the system and an ancillary system. The RHP measure \ts{quantifies} the non-CPTPness of the dynamical map $\mathcal{E}$ over the time evolution. It has been shown that the RHP measure is up to a constant factor equivalent to the decay rate measure 
\begin{equation}
R(t_1\rightarrow t_3)=\sum_k \int_{t_1} ^{t_3} \frac{|\gamma_k(t)|- \gamma_k(t)}{2}dt,
\label{Eq: decayrate}
\end{equation}
where $\gamma_k(t)$ is the time-dependent decay rate in the canonical form of the master equation~\cite{hall2014canonical}.

An alternative approach to quantify the quantum non-Markovianity is an increase of the distinguishability called Breuer, Laine and Piilo (BLP) measure~\cite{breuer2009measure,laine2010measure}, defined by
$\mathcal{N}_{\rm BLP}(t_1\rightarrow t_3) =\max_{\rho_1,\rho_2} \int_{\sigma>0}\sigma(t) dt$,
where $\sigma(t)= \frac{d}{dt}D(\rho_1(t),\rho_2(t))$ and $D(\rho_1,\rho_2)=\frac{1}{2}\|\rho_1-\rho_2\|_1$ is the trace distance between two quantum states $\rho_1,\rho_2$. 
Since CPTP maps never increase the trace distance, $\mathcal{N}_{\rm BLP}(t_1\rightarrow t_3)$ can witness the non-CPTPness during the evolution.

\emph{Subsystem frame.---}
In our work, we divide the whole Hilbert space into a logical subsystem, where we encode the quantum information for QIP, and a gauge subsystem. The concept of subsystem decomposition has been introduced in bosonic QEC, which separates the logical \ts{degrees of} freedom from the other gauge \ts{degrees of} freedom~\cite{pantaleoni2020modular,chamberland2022building,xu2023autonomous}. In our work, the gauge subsystem contains the information of syndrome measurement for QEC and that of  the Bell measurement for quantum teleportation. 
We introduce the subsystem isomorphism operator $V_S$ such that 
\begin{align}
V_{\rm S} U_m\ket{k}_{\rm C}=(Q_m\ket{k}_{\rm L})\otimes\ket{m}_{\rm G},
\label{Eq: subsystem}
\end{align}
where $\ket{k}_{\rm C}$ is a codeword, $
{U}_m$ is a unitary operator associated with the information of syndrome measurements or Bell measurements, and $Q_m$ is the unitary operator acting on the logical subsystem. Now, we refer to $\ket{k}_L$ and $\ket{m}_G$ as the logical and gauge subsystem states. The gauge subsystem behaves similarly to the environment, and the feedback operation based on the gauge state induces the backflow of the quantum information into the logical subsystem. In general, we are not always able to construct the subsystem isomorphism $V_S$ that transmits the information of the unitary $U_m$ to the gauge state $\ket{m}_G$. This generally results in a residual logical error that cannot be corrected under any feedback operations depending on the gauge information.

We first show an example of QEC under the Knill-Laflamme (KL) condition, and construct the subsystem isomorphism satisfying Eq.~\eqref{Eq: subsystem}.
Let the noise channel $\mathcal E(\cdot)=\sum_m F_m(\cdot)F_m^\dagger$ satisfy the KL condition $P_{\rm C}F_m^\dagger F_nP_{\rm C}=\delta_{mn}d_mP_{\rm C}$, where $P_{\rm C}$ is a projector onto the code space $\mathcal H_{\rm C}$.  
By polar decomposition, each $F_m P_{\rm C}$ can be written as $\sqrt{d_m}U_m P_{\rm C}$ with some unitary operator $U_m$.  
The subspaces $\mathcal H_m\coloneq{\rm Im}\,U_mP_{\rm C}U_m^\dagger$ are mutually orthogonal, and let $\{\ket{k,m}\}_k$ denote an orthonormal basis on $\mathcal H_m$.  
Since $U_m$ maps the codewords $\{\ket{k}_{\rm C}\}$ to $\mathcal H_m$ while preserving inner products, there exists a unitary matrix $\tilde Q_m$ such that $U_m\ket{k}_{\rm C}=\sum_l(\tilde Q_m)_{lk}\ket{l,m}$.
Defining orthogonal bases $\{\ket{k}_{\rm L}\}$ and $\{\ket{m}_{\rm G}\}$ for the logical and gauge spaces $\mathcal H_{\rm L}$ and $\mathcal H_{\rm G}$, respectively, we set  
$V_{\rm S}\ket{k,m} \coloneq \ket{k}_{\rm L}\otimes\ket{m}_{\rm G}$.  
Then, the composite action $V_{\rm S}U_m$ on the code space factorizes as Eq. \eqref{Eq: subsystem}, where $Q_m$ on $\mathcal H_{\rm L}$ is defined by $Q_m\ket{k}_{\rm L}=\sum_l(\tilde Q_m)_{lk}\ket{l}_{\rm L}$.  
This construction explicitly realizes the subsystem decomposition of the Hilbert space of the system into logical and gauge degrees of freedom, enabling a clear representation of the action of noise in this tensor-product form.
Refer to End Matter (EM) for details of the subsystem unitary isomorphism $V_{\rm S}$. In the case of cat and squeezed cat codes, a similar decomposition to separate the logical and gauge modes has been introduced~\cite{chamberland2022building,xu2023autonomous}.  For subsystem decomposition of quantum teleportation, refer to EM.

We also investigate the Pauli-based QEC under the KL condition for the explicit construction of $V_{\rm S}$. We encode the codewords through a Clifford unitary $U_{\mathrm{CL}}$ as $\ket{k}_{\rm C}=U_{\mathrm{CL}}\ket{k}_{\rm L} \otimes \ket{0}^{\otimes N_{\rm G}}_{\rm G}$, where $N_{\rm G}$ is the number of ancillary qubits for syndrome, or gauge qubits. Consider a stochastic Pauli noise with the error operator $F_m=\sqrt{d_m}P_{\rm e}^{(m)}$. Then, the error subspaces are given by ${\mathcal H}_m={\rm Im}\ P_e^{(m)} P_{\rm C} P_e^{(m)}$, which are orthogonal to each other. By choosing the subsystem isomorphism as $V_S=U_{\mathrm{CL}}^\dagger$, we can verify the consistency of the previous argument and construct an orthonormal basis $\{\ket{k,m}\}_k$ on $\mathcal H_m$ as $\ket{k,m}=U_{\mathrm{CL}} \ket{k}_{\rm L} \otimes \ket{m}_{\rm G}$.
A state in the error subspace ${\mathcal H}_m$ can be expressed in the subsystem frame as $V_{\rm S} P_{\rm e}^{(m)} U_{\mathrm{CL} }\ket{\psi}_{\rm L} \otimes \ket{0}^{\otimes N_{\rm G}}_{\rm G}= P_e^{\prime (m)} \ket{\psi}_{\rm L}  \otimes \ket{0}^{\otimes N_{\rm G}}_{\rm G}$, where $P^{\prime (m)}_e = U_{\mathrm{CL} }^\dagger P^{\prime (m)}_e U_{\mathrm{CL} }$ is again a Pauli operator. By denoting $P_e^{\prime (m)}= P_{\rm L}^{\prime (m)} \otimes \vec{P}_{\rm G}^{\prime (m)}$, we get $P_e^{\prime (m)} \ket{\psi}_{\rm L}  \otimes \ket{0}^{\otimes N_{\rm G}}=  P_{\rm L}^{\prime (m)}  \ket{\psi}_{\rm L}  \otimes \ket{m}_{\rm G} $ with $\ket{m}_{\rm G}= \vec{P}_{\rm G}^{\prime (m)} \ket{0}_{\rm G}^{\otimes N_{\rm G}}$.
The orthogonality of $\{\ket{m}_{\rm G}\}$ follows from the orthogonality in error subspaces $\{\mathcal H_m\}$. Thus we can identify as $Q_m=  P_{\rm L}^{\prime (m)}$ for Pauli stabilizer codes.

\emph{Non-Markovianity in QIP---}
We describe how the non-Markovian processes emerge in quantum information processing. In the subsystem frame described by the subsystem unitary operator $V_{\rm S}$, we define the logical state by $\rho_{L,1}= \mathrm{Tr}_{\rm G}[V_{\rm S} \rho_{\rm in} V_{\rm S}^\dag]$ for the \ts{initial} quantum state in the total composite system $\rho_{\rm in}$ \ts{at time $t_1$}. We also define the process under the subsystem frame:
\begin{equation}
\mathcal{E}^{\rm L}_{t_1\rightarrow t} (\rho_{{\rm L},1}) = \mathrm{Tr}_{\rm G}[ V_{\rm S} \mathcal{F}_{t_1\rightarrow t}(\rho_{\rm in}) V_{\rm S}^\dag ],
\end{equation}
where $\mathcal{F}_{t_1,t}$ is the superoperator for time evolution from time $t_1$ to $t$. Note that the time evolution from $t$ to $s$ 
\begin{equation}
\mathcal{E}^{\rm L}_{t\rightarrow s}= \mathcal{E}^{\rm L}_{t_1\rightarrow  s}\circ (\mathcal{E}_{t_1\rightarrow  t}^{\rm L})^{-1}
\end{equation} for $s>t$  can be a non-CPTP map, which indicates the presence of non-Markovian processes. We can quantify the amount of non-Markovianity through non-Markovian measures such as \ts{the} RHP and BLP measures~\cite{rivas2010entanglement,breuer2009measure}. Now, we illustrate the non-Markovian processes in quantum information processing. We discuss non-Markovianity in QEC and quantum teleportation in the main text and EM, respectively.

\emph{Non-Markovianity in QEC under KL condition.---}
Suppose that the quantum state on the code subspace $\rho_{\rm C}$ is affected by a noise process $\mathcal{F}$ under the KL condition. 
Then, the state is changed into $\rho'_{\rm C}=\sum_m d_m U_m \rho_{\rm C} U_m^\dag$. 
Here, $d_m$ is the probability that the state is measured in the erroneous subspace ${\mathcal H}_m$. 
 Under the subsystem frame in Eq.~\eqref{Eq: subsystem}, we transform the noisy state $\rho'_{\rm C}$ to $V_{\rm S} \rho'_{\rm C} V_{\rm S}^\dag= \sum_m d_m Q_m \rho_{\rm L} Q_m^\dag \otimes \ket{m}_{\rm G}\bra{m}_{\rm G}$. 
 By partially tracing out the gauge subsystem, we have
$\mathrm{Tr}_{\rm G}[V_{\rm S} \rho'_{\rm C} V_{\rm S}^\dag ] = \sum_m d_m  Q_m \rho_{\rm L} Q_m^\dag=:\mathcal{E}^{\rm L}_{t_1 \rightarrow t_2}[\rho_{\rm L}]$,
which indicates that the logical subsystem state is decohered due to the noise map $\mathcal{E}^{\rm L}_{t_1 \rightarrow t_2}$. 
The QEC operation $\mathcal{R}$ can correct the noisy state $\mathcal{E}^{\rm L}_{t_1 \rightarrow t_2}[\rho_{\rm L}]$ into the noiseless state $\rho_{\rm L}$, where
$\mathcal{R}(\cdot)=\sum_m R_m (\cdot) R_m^\dag$, and
$R_m = Q_m^\dag \otimes \ket{0}_{\rm G}\bra{m}_{\rm G}$.
This implies that the logical subsystem state is perfectly recovered $\mathcal{E}^{\rm L}_{t_1 \rightarrow t_3}=\mathcal{I}$ and then the QEC operation on the gauge subsystem can be regarded as a non-Markovian process $\mathcal{E}^{\rm L}_{t_2 \rightarrow t_3}=(\mathcal{E}^{\rm L}_{t_1 \rightarrow t_2})^{-1}$.

\emph{Non-Markovianity in the three-qubit code.---} Here, we first show the example of the three-qubit code~\cite{nielsen2010quantum} under the bit-flip error $\mathcal{F}= \mathcal{F}^{(1)}_{p}\circ \mathcal{F}^{(2)}_{p}\circ \mathcal{F}^{(3)}_{p}$, defined by
$\mathcal{F}^{(i)}_{p}[\rho]=(1-p)\rho +p X_i\rho X_i$, and we choose $V_{\rm S}={\rm CNOT}_{1,2}{\rm CNOT}_{2,3}$ from the encoding unitary of three-qubit code.  We have $\mathcal{E}^{\rm L}_{1\rightarrow 3}(\rho_{L,1})=
\mathrm{Tr}_{23}[V_{\rm S} \mathcal{R}\circ\mathcal{F}(\rho_{\rm C}) V_{\rm S}^\dag]= \mathcal{F}^{(1)}_{p^2(3-2p)}(\rho_{L,1})$ and $\mathcal{E}^{\rm L}_{1\rightarrow 2}(\rho_{L,1})=\mathrm{Tr}_{23}[V_{\rm S} \mathcal{F}(\rho_{\rm C}) V_{\rm S}^\dag]=\mathcal{F}^{(1)}_{p}(\rho_{L,1})$ when using the straightforward majority vote strategy.
Then we have
$\mathcal{E}^{\rm L}_{2\rightarrow 3}=\mathcal{E}^{\rm L}_{1\rightarrow 3}\circ (\mathcal{E}^{\rm L}_{1\rightarrow 2})^{-1}
    =\mathcal{F}^{(1)}_{-p(1-p)}$
is non-CPTP for any $p\in(0,1/2)$. Therefore, the error-corrected map can be regarded as a non-Markovian process.

Now, we also discuss the continuous QEC model and derive the canonical form of the master equation for the logical state:
$\frac{d}{dt}\rho_{\rm L}(t)= \frac{f'(t)}{1-2f(t)} (X \rho_{\rm L}(t) X- \rho_{\rm L}(t))$,
where $f(t)=(1- e^{-\gamma t}) p^2(3-2p)+e^{-\gamma t}p$ (see SM for derivation~\cite{Supplementary_Materials}).
Note that $f'(t) <0$  holds for all $t>0$ when $p<1/2$, which indicates that the continuous error correction procedures induce the non-Markovian effect during the whole time evolution. The decay rate measure in Eq. \eqref{Eq: decayrate} for the three-qubit code when reducing the error from $p$ to $q$ reads 
\begin{equation}
R_{p \rightarrow q}= \frac{1}{2} \mathrm{log}\bigg( \frac{1- 2q}{1-2p} \bigg).
\end{equation}
We will later discuss how this decay rate measure lowers the QEM sampling cost. 

\begin{figure}[htbp]
    \centering
    \includegraphics[width=0.85\linewidth]{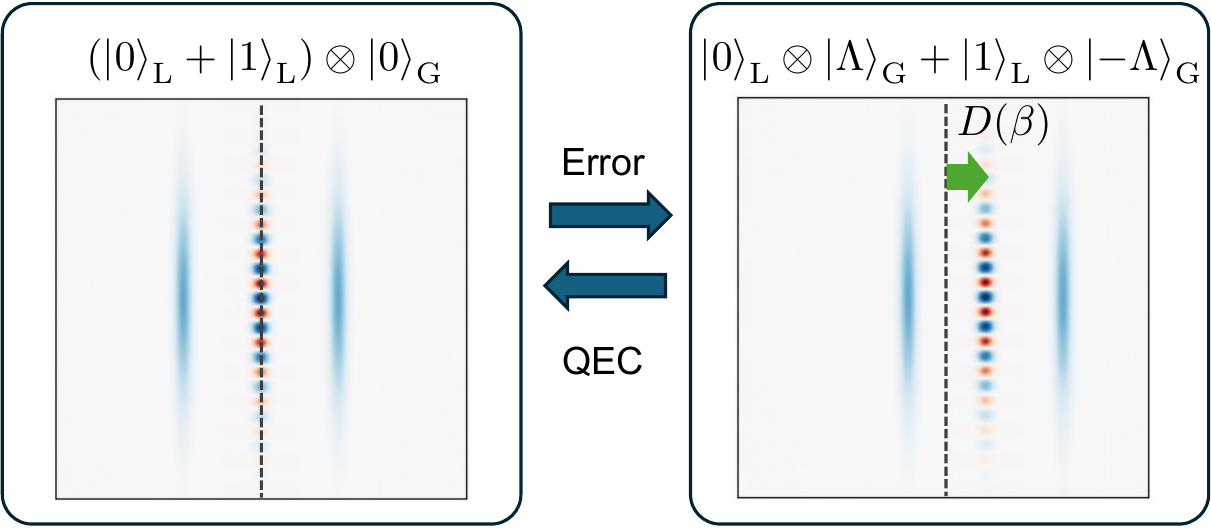}
    \caption{Displacement error and QEC process in physical frame and subsystem frame for the squeezed cat code.
    }
    \label{Fig: sqcatQEC}
\end{figure}

\emph{Non-Markovianity in bosonic QEC.---}
We discuss the non-Markovian dynamics present in error correction procedures in cat and squeezed cat codes. Since cat codes are a special case of squeezed cat codes, we mainly illustrate our result with squeezed cat codes. We here consider a displacement error. Ref.~\cite{xu2023autonomous} introduces the subsystem unitary isomorphism $V'_{\rm S}$ satisfying $V'_{\rm S} a V_{\rm S}'^\dag  
= Z_{\rm L} \otimes (\alpha e^{-r} I_{\rm G} + \mathrm{cosh}(r) a_{\rm G}-\mathrm{sinh}(r) a_{\rm G}^\dag )+O(e^{-2\alpha^2})$. Then, the displacement error $D(\beta)$ for a complex number $\beta \in \mathbb{C}$ in the subsystem frame reads
$V'_{\rm S} D(\beta) V_{\rm S}^{\prime \dag} \sim e^{2 i \alpha \mathrm{Im}(\beta) Z_{\rm L}\otimes I_{\rm G}} \times e^{Z_{\rm L} \otimes (\Lambda a_{\rm G}^\dag-\Lambda^* a_{\rm G})}$, where $\Lambda= \mathrm{Re}(\beta)e^r +i \mathrm{Im}(\beta)e^{-r}$. While $e^{2 i \alpha \mathrm{Im}(\beta) Z_{\rm L}\otimes I_{\rm G}} $ purely works as a logical operation, $e^{Z_{\rm L} \otimes (\Lambda a_{\rm G}^\dag-\Lambda^* a_{\rm G})}$ induces an entangling operation between the logical and the gauge modes, hence resulting in the dephasing error for the logical state. See SM for details. For $\beta>0$ and the initial state $\ket{\psi_0}=(c_0 \ket{0}_{\rm L}+c_1 \ket{1}_{\rm L})\otimes \ket{0}_{\rm G}$, the displacement operation purely induces the dephasing error to the quantum state: 
$\ket{\psi_\Lambda} = c_0 \ket{0}_{\rm L} \otimes \ket{\Lambda}_{\rm G}+c_1  \ket{1}_{\rm L} \otimes \ket{-\Lambda}_{\rm G}$.
We illustrate the displacement error and QEC  both in physical frame and subsystem frame in Fig.~\ref{Fig: sqcatQEC}.

Now, the dissipative QEC process $\frac{d}{dt}\rho_{\rm LG}(t)=\frac{\gamma}{2}(2 L_{d} \rho_{\rm LG}(t) L_{d}^\dag - L_{d}^\dag L_d \rho_{\rm LG}(t) - \rho_{\rm LG}(t)  L_{d}^\dag L_d )$ to 
$\ket{\psi_\Lambda}$
with $L_d=Z_{\rm L} \otimes a_{\rm G}$ yields the canonical form of the master equation as follows:
$\frac{d}{dt} \rho_{\rm L}(t)= -\Lambda^2 \gamma e^{-\gamma t} (Z_{\rm L} \rho_{\rm L}(t) Z_{\rm L} - \rho_{\rm L}(t))$,
which indicates that the \ts{off}-diagonal term rephases due to the dissipative QEC processes.
Refer to SM for derivation~\cite{Supplementary_Materials}, and we show an excellent agreement 
with the numerical simulations in EM. Note that the decay rate measure reads $R(0\rightarrow T)= \Lambda^2 \gamma \int_0 ^T e^{-\gamma t}dt = \Lambda^2 (1-e^{-\gamma T})$ during the time $t\in[0,T]$. We will later discuss how this decay rate measure leads to the reduction of the QEM sampling cost overhead.

\emph{Interplay with QEM.---}
Quantum error mitigation is a series of hardware-efficient error suppression techniques that estimate noiseless results via post-processing of measurement outcomes~\cite{temme2017error,endo2018practical,endo2021hybrid,cai2023quantum}. The relationship between non-Markovian processes and QEM has been pointed out in Ref. \cite{hakoshima2021relationship}. This work demonstrates that the cost of a QEM, specifically stochastic QEM, decreases due to the negativity of the noise process. Later, information-theoretic analysis for QEM via distance measures has been introduced, which shows the inevitable exponentially increasing sampling cost of QEM~\cite{takagi2022fundamental,takagi2023universal,tsubouchi2023universal,quek2024exponentially}. Note that the sampling overhead of QEM can be lower-bounded by the inverse of the distance measures of noisy quantum states. For example, Ref. \cite{takagi2022fundamental} shows that the sufficient number of samples $M$ of the $(Q,K)$-mitigation protocol need be:
\begin{equation}
M \geq \mathrm{max}_{\rho, \sigma} \bigg[\frac{D(\rho, \sigma) - 2 b_{\rm max}}{D(\tilde{\rho}^{(K)}_Q, \tilde{\sigma}^{(K)}_Q)}  \bigg]^2 \frac{\log\big(2/\varepsilon \big)}{2 \delta^2},
\label{Eq:flimit}
\end{equation}
where $\rho$ and $\sigma$ are the target noiseless quantum states, $b_{\rm max}$ is the maximum bias for the error-mitigated outcome, $\tilde{\rho}_Q^{(K)}= \bigotimes_{k=1}^K \bigotimes_{q=1}^Q [\mathcal{E}_q^k(\rho)]$, and $\tilde{\sigma}_Q^{(K)}= \bigotimes_{k=1}^K \bigotimes_{q=1}^Q [\mathcal{E}_q^k(\sigma)]$ for noise processes $\mathcal{E}_q^k$. For $ \varepsilon$ and $\delta$, we require that the $(Q,K)$-mitigation protocol realizes the bias smaller than $b_{\rm max} + \delta$ with the success probability $1-\varepsilon$. Refer to \cite{takagi2022fundamental} for detailed definitions of $(Q,K)$-mitigation protocols.

Note that while Markovian dynamics necessarily render the distinguishability to decrease, non-Markovian dynamics can increase distinguishability, witnessing the presence of non-Markovian processes~\cite{breuer2009measure}. Based on our results, QEC generally induces non-Markovianity for rephasing quantum states and can increase the trace distance. Therefore, it is clear that the necessary sample number $M$ decreases due to the non-Markovian dynamics induced by QEC, which highlights the practical significance of integrating QEC with QEM.

Let us consider a single qubit case under dephasing noise $\mathcal{E}_p(\cdot)=(1-p) (\cdot)+  pZ (\cdot) Z $. For unbiased QEM, we have $b_{\rm max}=0$ and the right-hand side of Eq.~\eqref{Eq:flimit} reads~\cite{takagi2022fundamental}  
$M_p =   \frac{\log(2/\varepsilon )}{2 \delta^2(1- 2p)^2}$. 
Now, suppose that QEC, due to the continuous feedback model, reduces the error rate from $p$ to $q$. The QEM for the error rate $q$ incurs $M_q=   \frac{\log(2/\varepsilon )}{2 \delta^2(1- 2q)^2}$, which gives 
\begin{equation}
M_q = M_p~ \mathrm{exp}[-4 R_{p \rightarrow q}].
\end{equation}
This can be interpreted as non-Markovianity due to QEC reducing the sampling overhead. For the squeezed cat code, denoting the sampling overhead for QEM to obtain the unbiased estimator at time $t$ in the canonical master equation as $M(t)$, we obtain $M(T)=M(0) \mathrm{exp}[-4 R(0\rightarrow T)]$.

\emph{Conclusion and Discussions.---} 
In this work, we reveal the non-Markovianity in QIP. By introducing the subsystem unitary operator, we show that the feedback operation based on the measurement result generates the information flow into the logical subsystem, leading to the negativity of the dynamical maps. We explore QEC and quantum teleportation as concrete examples. We further bridge this argument with the scenario of QEM, clearly showing that the negativity due to QEC contributes to reducing the sampling overhead of QEM. 

Our subsystem unitary isomorphism may offer an efficient strategy for the combination of QEC and QEM~\cite{suzuki2022quantum,xiong2020sampling,liu2025virtual,araki2025correcting,zhang2025demonstrating,wahl2023zero,piveteau2021error,lostaglio2021error}. For example, under the subsystem frame under the unitary $V_{\rm S}=\mathrm{CNOT}_{1,2} \mathrm{CNOT}_{2,3}$, the bit flip error in the second qubit only leads to the error in the gauge qubit; thus, if the reset operation of gauge qubits is straightforward, we only need to reset the gauge qubit for QEC. In this case, we should not rely on QEM for suppressing gauge qubit errors because it incurs a sampling overhead.

Finally, while we derive the dynamical map and the master equation for time evolution and point out the negativity for specific examples, we can consider other instances. For example, the subsystem decomposition for bosonic codes was first proposed for GKP codes~\cite{pantaleoni2020modular}. While it may be complicated to derive analytical dynamical evolution for logical mode in GKP codes during QEC procedures, it is natural to observe a non-Markovian negative map. In addition, more complicated QEC, e.g., surface code QEC~\cite{fowler2012surface}, should involve non-Markovianity, and it may be interesting to numerically simulate non-Markovianity in such practical codes.  

\emph{Acknowledgments.---} 
We acknowledge useful discussions with Keisuke Fujii, Hiroyasu Tajima, Yuichiro Matsuzaki, and Hiroyuki Harada. This work was supported by JST [Moonshot R\&D] Grant No.~JPMJMS2061; MEXT Q-LEAP, Grant No.~JPMXS0120319794; JST CREST Grant No.~JPMJCR23I4, and No.~JPMJCR25I4; JSPS KAKENHI Grant No. JP24K16979. 
We perform all the numerical calculations with QuTiP \cite{qutip5}.

\bibliographystyle{apsrev4-1}
\bibliography{bib}


\clearpage


\newpage
\section{End Matter}

\emph{Quantum teleportation under subsystem frame.---} 
We can also discuss quantum teleportation~\cite{bennett1993teleporting,pirandola2015advances} under the subsystem frame. Suppose that Alice aims to send the quantum information $\ket{\psi}_{A_1}$ to Bob using the shared Bell state $\ket{\Phi_{00}}_{A_2 B}=\frac{1}{\sqrt{2}}(\ket{00}_{A_2 B}+\ket{11}_{A_2 B})$ and the adaptive operation after the Bell measurement \ts{over $A_1A_2$ performed by Alice}. Denoting the four types of Bell states $\{\ket{\Phi_{m_1m_2}} \}_{m_1,m_2=0}^1$, where we define $\ket{\Phi_{m_1m_2}} =X_{A_2}^{m_1}Z_{A_2}^{m_2} \ket{\Phi_{00}}_{A_2 B}$, we define the subsystem unitary operator
\begin{equation}
V_{\rm S}= \sum_{m_1,m_2=0}^1 \ket{m_1m_2}_{A_1 A_2}\bra{\Phi_{m_1m_2}}_{A_1 A_2}, 
\end{equation}
and the adaptive operations are represented as $R_{m_1m_2}= \ket{00}_{A_1 A_2}\bra{m_1m_2}_{A_1 A_2}\otimes Q_{m_1m_2}$ with $Q_{m_1m_2}= X_B^{m_1} Z_B^{m_2}$. Note that $V_{\rm S}$ can be performed by a controlled-NOT operation followed by a Hadamard gate. As is well known, before the adaptive operation with the Bell measurement, Bob's state is initially a completely mixed state because the composite state is written as
\begin{equation}
V_{\rm S}\ket{\psi}_{A_1} \otimes \ket{\Phi_{00}}_{A_2 B} = \frac{1}{2} \sum_{m_1,m_2=0}^1 \ket{m_1m_2}_{A_1 A_2} \otimes Q_{m_1m_2} \ket{\psi}_B.
\end{equation}
Partially tracing the freedom of Alice gives a completely mixed state in Bob. 
With the adaptive operation, we can decouple the freedom of $B$ from $A$ with the teleportation being completed. Regarding the subsystem $B$ and $A_1 A_2$ to be the logical and gauge subsystems $L$ and $G$, we have similar arguments to QEC cases. We emphasize that these arguments also apply to other teleportation schemes, such as magic state teleportation.

\emph{Non-Markovianity in Quantum teleportation.---} 
We consider the continuous measurement and feedback model for quantum teleportation. We first introduce the continuous feedback model described by the GKSL master equation with the dissipators $\{ R_m\}_m$ as
$\frac{d}{dt} \rho_{\rm LG}(t) = \gamma  \left[ \left(\sum_m R_m \rho_{\rm LG}(t) R_m ^\dag\right) - \rho_{\rm LG}(t) \right]$,
where $\rho_{\rm LG}(t)$ is the state under the subsystem frame, $\gamma$ is a positive constant, and we set $m=(m_1, m_2)$ and $R_m=\ket{00}\bra{m_1m_2}\otimes Q_{m_1 m_2}^\dag$. When tracing out the gauge subsystem, 
 the master equation
 is reduced to 
$\frac{d}{dt} \rho_{\rm L}(t) = \gamma (\ket{\psi}_{\rm L}\bra{\psi}_{\rm L}-  \rho_{\rm L}(t))$,
which can be easily solved to give a non-Markovian process $\rho_{\rm L}(t) = (1- e^{-\gamma t} )\ket{\psi}_{\rm L}\bra{\psi}_{\rm L}+e^{-\gamma t} \rho_{\rm L}(0)$ (see SM for derivation~\cite{Supplementary_Materials}). While Ref. \cite{tserkis2025information} discusses the information backflow in quantum teleportation, we derive the continuous dynamics for quantum teleportation with negative decay rates. For $t>0$, we can show that the canonical master equation can be rewritten as
\begin{equation}
\begin{aligned}
\frac{d \rho_{\rm L}(t)}{dt} &= - \frac{\gamma e^{-\gamma t}}{4(1-e^{-\gamma t})}\sum_{k=1}^3 \mathcal{L}[P_k](\rho_{\rm L}(t)), \\
\label{Eq: teleportation}
\end{aligned}
\end{equation}
where $P_1=X$, $P_2=Y$ and $P_3=Z$ with $\mathcal{L}[P_k](\rho) = P_k \rho P_k - \rho$ with $- \frac{\gamma e^{-\gamma t}}{4(1-e^{-\gamma t})}<0$. See SM for detailed derivations.
We cannot use the canonical master equation at $t=0$ in Eq.~\eqref{Eq: teleportation}
, since the right-hand side diverges. While the non-Markovianity effect is explained by information backflow from the environment, our result clearly shows that the information classically sent by Alice imparts information flow to Bob, leading to the reconstruction of the quantum state. However, because the initial state of Bob is a completely mixed state, we cannot construct the dynamical map from $t=0$ to $t=\delta t$ for the infinitesimal time step $\delta t$. While we cannot compute the decay rate measure in the total teleportation procedure due to the singularity in $t=0$, the decay rate measure during $t=\Delta t$ to $t=T$ reads:
\begin{equation}
\int_{\Delta t} ^T\frac{\gamma e^{-\gamma t}}{4(1-e^{-\gamma t})} dt = \mathrm{log}\bigg[\frac{1-e^{-\gamma T} }{1-e^{-\gamma \Delta t}} \bigg]. 
\end{equation}

\begin{figure}[h!]
    \centering
\includegraphics[width=\linewidth]{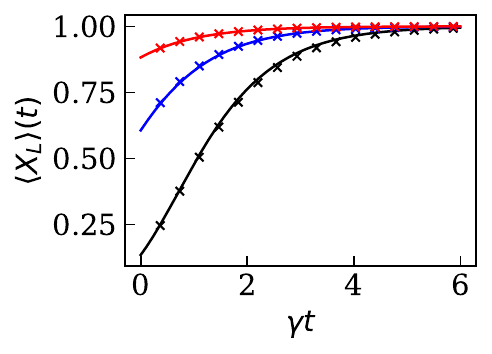}
    \caption{Dynamics of the QEC process in the squeezed cat code against the displacement error.
    The expectation value of the logical Pauli $X$ operator is plotted. The solid lines correspond to results from the non-Markovian master equation, 
     while the simulation results of the QEC circuit proposed in Ref.~\cite{shitara2025exploiting} are plotted with crosses. The displacement amplitude is set to be $\Lambda=$1 (red), 0.5 (blue), and 0.25 (red). Other parameters are set to be $\alpha=2$ and $ r=1.3$.
    }
    \label{Fig: sqcat}
\end{figure}

\emph{Validity of the effective non-Markovian master equation for the squeezed cat QEC---} To demonstrate the validity of the effective non-Markovian master equation for the logical space, we perform numerical simulations. 
Suppose that we initially prepare a logical state $\ket{+}_{\rm L}\otimes\ket{0}_{\rm G}$, and it undergoes a displacement error $\Lambda$ in the real direction.
Then the master equation for the logical state can be analytically solved, and we see that the QEC process recovers the off-diagonal component, or coherence, which results in the expectation value of the logical $X$ operator as
    $\ev{X_{\rm L}}(t)=\Tr[\rho_{\rm L}(t)X_{\rm L}]=e^{-2\Lambda^2 e^{-\gamma t}}$. 
On the other hand, Ref.~\cite{shitara2025exploiting} proposes an autonomous QEC circuit that approximately realizes the dissipative process by the dissipator $L_d$ through a repetitive interaction with an ancillary qubit and the reset of the qubit.
One cycle of the QEC circuit results in the dissipative process with $\gamma t = 2\pi^2e^{-r}/\alpha^2$.
We apply the QEC circuits multiple times to dissipate the excitations in the gauge mode.
In Fig.~\ref{Fig: sqcat}, we compare the analytical result of $\ev{X_{\rm L}}(t)$
and the result from the numerical simulation of the QEC circuit.
They show an excellent agreement, which validates the effective description of the non-Markovian master equation in the logical space.

\emph{Subsystem unitary isomorphism.---} 
We introduce a unitary isomorphism $V_{\rm S}$ that relates the system Hilbert space with the tensor product of the logical and gauge spaces via subsystem decomposition.
Let $\mathcal{H}_{\rm C}(\subset \mathcal{H}_{\rm S})$ and $P_{\rm C}$ be the code space and the projector onto it, respectively.
We denote the orthonormal basis on $\mathcal{H}_{\rm C}$ (i.e., codewords) by $\{ \ket{k}_{\rm C} \}_{k=1}^K$.
Suppose that the noise process $\mathcal E$ has a Kraus representation
\begin{align}
    \mathcal{E} (\cdot)=\sum_{m=1}^M F_m \cdot F_m^\dagger.
\end{align}
When the Knill-Laflamme condition is satisfied, the Kraus operators $\{F_m\}_{m=1}^M$ can be chosen so that
\begin{align}
    P_{\rm C} F_m^\dagger F_n P_{\rm C} = \delta_{mn} d_m P_{\rm C},
\end{align}
where $d_m\ge 0$ and $\delta_{mn}$ is the Kronecker delta defined by 
\begin{align}
    \delta_{mn} \coloneq
    \begin{cases}
        1  & (m=n),\\
        0  & (m\neq n).
    \end{cases}
\end{align}
By the polar decomposition of $F_m P$, we obtain a unitary operator $U_m$ satisfying
\begin{align}
    F_m P_{\rm C} = U_m \cdot \sqrt{P_{\rm C} F_m^\dagger F_m P_{\rm C}} =\sqrt{d_m}U_m P_{\rm C}.
\end{align}
We define a projector onto the erroneous subspace induced by $F_m$ by $P_{\rm C}^m \coloneq U_mP_{\rm C}U_m^\dagger$. The projectors $\{ P_{\rm C}^m \}_{m=1}^M$ are mutually orthogonal as
\begin{align}
    P_{\rm C}^mP_{\rm C}^n &= U_mP_{\rm C}U_m^\dagger U_nP_{\rm C}U_n^\dagger\\
    &= U_m \cdot \frac{1}{\sqrt{d_m}} P_{\rm C} F_m^\dagger \cdot \frac{1}{\sqrt{d_n}} F_n P_{\rm C} \cdot U_n^\dagger\\
    &= \delta_{mn} P_{\rm C}^m.
\end{align}
Let $\mathcal{H}_m \coloneq {\rm Im}\ P_{\rm C}^m$ be the Hilbert space corresponding to the projector $P_{\rm C}^m$.
Let $\{\ket{k,m}\}_{k=1}^K$ be an orthonoraml basis on $\mathcal{H}_m$.
Then, $\{\ket{k,m}\}_{k,m=1}^{K,M}$ is also an orthonormal basis on $\mathcal{H}_{\rm S}' \coloneq \oplus_{m=1}^M\mathcal{H}_m$, since $\mathcal{H}_m$'s are mutually orthogonal.
Noting that $U_m$ maps a vector in the code space to that in $\mathcal{H}_m$ and keeps inner product, there exists a $K\times K$ unitary matrix $(\tilde{Q}_m)_{kl}$ that satisfies
\begin{align}
    U_m \ket{k}_{\rm C} =\sum_{l=1}^K (\tilde{Q}_m)_{lk} \ket{l,m}.
\end{align}

To decompose $\mathcal{H}_{\rm S}'$ into logical and gauge subsystems, we define two Hilbert spaces $\mathcal{H}_{\rm L}$ and $\mathcal{H}_{\rm G}$ with orthogonal bases $\{\ket{k}_{\rm L}\}_{k=1}^K$ and $\{\ket{m}_{\rm G}\}_{m=1}^M$, respectively.
We define a unitary isomorphism $V_{\rm S}$ from $\mathcal{H}'_{\rm S}$ to $\mathcal{H}_{\rm L}\otimes\mathcal{H}_{\rm G}$ by 
\begin{align}
    V_{\rm S} \ket{k,m} \coloneq \ket{k}_{\rm L}\otimes\ket{m}_{\rm G}.
\end{align}
Then, the action of $V_{\rm S}U_m$ on the code space can be written as 
\begin{align}
    V_{\rm S}U_m \ket{k}_{\rm C} &= \sum_{l=1} (\tilde{Q}_m)_{lk} \ket{l}_{\rm L}\otimes\ket{m}_{\rm G} \nonumber\\
    &= (Q_m \ket{k}_{\rm L})\otimes\ket{m}_{\rm G},
\end{align}
where we defined an unitary operator $Q_m$ on $\mathcal{H}_{\rm L}$ by
\begin{align}
    Q_m \ket{k}_{\rm L} \coloneq \sum_{l=1}^K (\tilde{Q}_m)_{lk} \ket{l}_{\rm L}.
\end{align}

Some remarks are in order.
First, when $\dim \mathcal{H}_{\rm S}=KM$ and hence $\mathcal{H}_{\rm S}=\mathcal{H}'_{\rm S}$, then $V_{\rm S}$ is a unitary isomorphism between $\mathcal{H}_{\rm S}$ and $\mathcal{H}_{\rm L}\otimes\mathcal{H}_{\rm G}$. 
Second, although the choice of the orthogonal basis $\{\ket{k,m}\}_{k=1}^K$ on $\mathcal{H}_m$ is arbitrary, it determines the explicit expression of the unitary matrix $\tilde{Q}_m$, and hence the action of $Q_m$ on the logical space.
In the situations discussed in the main text, there exists a physically natural choice.
When $\mathcal{H}_{\rm S}=\mathcal{H}'_{\rm S}$ and we identify $\mathcal{H}_{\rm S}$ with $\mathcal{H}_{\rm L}\otimes\mathcal{H}_{\rm G}$ under such a natural correspondence as in the Pauli-based QEC, we simply refer to $V_{\rm S}$ as a subsystem unitary operator. Note that, when expanding $\ket{k,m}$ in the computational basis representing the physical state of the qubits, $V_{\rm S}$ has a unitary matrix representation; therefore, $V_{\rm S}$ can be performed in a qubit-based quantum computer as a unitary operator. More specifically, we choose a Clifford decoding unitary $U_{\rm enc}^\dag$ as $V_{\rm S}$. In Table \ref{table: correspondence}, we exemplify the case of the three-qubit code. We can assign the bases as in the Table \ref{table: correspondence}, which corresponds to the choice of $V_{\rm S}=\mathrm{CNOT}_{1,2} \mathrm{CNOT}_{2,3}$. We assign the syndrome values of $Z_1Z_2$ and $Z_2Z_3$ to the label of the gauge subsystem $m=(m_1,m_2)$.

\begin{table}[h!]
    \centering
        \caption{Correspondence between $\ket{k,m}$ and $V_{\rm S}\ket{k,m}=\ket{k}_{\rm L} \otimes \ket{m}_{\rm G}$ and the syndrome values for the stabilizers.}
        \begingroup
        \renewcommand{\arraystretch}{1.2}
    \begin{tabular}{|c|c|c|}
    \hline  
    $\ket{k,m}$ &  $V_{\rm S}\ket{k,m}=\ket{k}_{\rm L} \otimes \ket{m}_{\rm G}$ & \{ $Z_1 Z_2$, $Z_2 Z_3$ \}  \\ \hline 
    $\ket{000}$ & $\ket{0}_{\rm L}\otimes \ket{00}_{\rm G}$ & \{0,0\} \\ \hline
    $\ket{001}$ &  $\ket{0}_{\rm L}\otimes \ket{01}_{\rm G}$ & \{0,1\} \\ \hline
    $\ket{010}$ & $\ket{0}_{\rm L}\otimes \ket{11}_{\rm G}$ & \{1,1\}\\\hline
    $\ket{100}$ & $\ket{1}_{\rm L}\otimes \ket{10}_{\rm G}$ & \{1,0\} \\ \hline
    $\ket{111}$ & $\ket{1}_{\rm L}\otimes \ket{00}_{\rm G}$ & \{0,0\} \\ \hline
    $\ket{110}$ & $\ket{1}_{\rm L}\otimes \ket{01}_{\rm G}$ & \{0,1\}  \\ \hline
    $\ket{101}$ & $\ket{1}_{\rm L}\otimes \ket{11}_{\rm G}$ & \{1,1\} \\ \hline
    $\ket{011}$ & $\ket{0}_{\rm L}\otimes \ket{10}_{\rm G}$ & \{1,0\} \\ \hline
    \end{tabular}
    \endgroup
    \label{table: correspondence}
\end{table}

\onecolumngrid
\clearpage
\begin{center}
	\Large
	\textbf{Supplementary Materials for: Non-Markovianity in Quantum Information Processing: \\
Interplay with Quantum Error Mitigation}
\end{center}

\setcounter{section}{0}
\setcounter{equation}{0}
\setcounter{figure}{0}
\setcounter{table}{0}
\renewcommand{\thesection}{S\arabic{section}}
\renewcommand{\theequation}{S\arabic{equation}}
\renewcommand{\thefigure}{S\arabic{figure}}
\renewcommand{\thetable}{S\arabic{table}}

\section{Subsystem decomposition for bosonic QEC} 
We discuss the cat and the squeezed cat code as a useful example of bosonic QEC. The subsystem decomposition for squeezed cat codes is an example of the subsystem frame representation~\cite{xu2023autonomous,chamberland2022building}. Note that cat codes are a specific example of the squeezed cat codes with the squeezing level $r=0$. We first discuss the cat code. We introduce the basis superposition states of the displaced Fock states:
\begin{equation}
\ket{\Phi_{n, \pm}}= \frac{1}{\sqrt{\mathcal{N}_{n,\pm}}} (D(\alpha) \pm (-1)^n D(-\alpha)) \ket{n},
\end{equation}
\ts{where $\mathcal{N}_{n,\pm}$ is the normalization factor.}
Here, $D(\alpha)=e^{\alpha a^\dag -\alpha^* a}$ is \ts{the} displacement operator \ts{with} an amplitude $\alpha \in \mathbb{C}$ and $\ket{n}$ is the Fock state with a non-negative integer $n \in \mathbb{N}_0$. Note that $n=0$ corresponds to the codewords of the cat states. While this basis set is not orthogonal, we can apply the Gram-Schmidt \ts{orthonormalization} procedure to obtain the \ts{orthonormal} set $\ket{\Phi'_{n, \pm}}$. We consider the transformation from the physical basis to the subsystem basis, denoted as \ts{$\{\ket{s}_{\rm L} \otimes \ket{n}_{\rm G}\}_{s=\pm,n\in\mathbb N_0}$}:
\begin{equation}
V_{\rm S} = \sum_{s=\pm, n=0} \ket{s}_{\rm L} \otimes \ket{n}_{\rm G} \bra{\Phi'_{n,s}}.
\end{equation}
In this case, because $V_{\rm S}$ maps the quantum state to a different Hilbert space, $V_{\rm S}$ is a unitary isomorphism operator. Now, in the subsystem frame defined by $V_{\rm S}$, we can show for an annihilation operator
\begin{equation}
V_{\rm S} a V_{\rm S}^\dag = Z_{\rm L} \otimes (\alpha I_{\rm G} + a_{\rm G} )+O(e^{-2 \alpha^2}),
\end{equation}
where $Z_{\rm L} \ket{\pm}_{\rm L} = \ket{\mp}_{\rm L}$ and $a_{\rm G}\ket{n}_{\rm G} = \sqrt{n} \ket{n-1}_{\rm G}$ for $n \geq 1$ and $a_{\rm G}\ket{0}_{\rm G}=0 $. This representation indicates that a photon loss error induces a \ts{logical} phase flip error with the gauge subsystem being affected by $\alpha I_{\rm G} + a_{\rm G} $. Since $n_{\rm G}=0$ for \ts{the} cat code, the gauge subsystem is invariant under the photon loss; therefore, we cannot detect the photon loss error. \ts{
To overcome this problem in the squeezed cat state, we consider the squeezed cat state defined by 
$S(r)\ket{\Phi_{0,\pm}}$, where $S(r)=e^{1/2(r^* a^2-r a^{\dag 2} )} ~(r>0)$ is the squeezing operator.
Then, a natural basis is modified to the squeezed displaced Fock state $S(r)\ket{\Phi'_{n,\pm}}$.
The subsystem isomorphism is also changed to $V'_{\rm S}=V_{\rm S} S(r)^\dagger$.
Accordingly, the annihilation operator in this subsystem frame is
\begin{align}
&V'_{\rm S} a V_{\rm S}'^\dag = V_{\rm S} S(r)^\dagger a S(r) V_{\rm S}^\dag \nonumber\\
=& Z_{\rm L} \otimes (\alpha e^{-r} I_{\rm G} + \mathrm{cosh}(r) a_{\rm G}-\mathrm{sinh}(r) a_{\rm G}^\dag )+O(e^{-2\alpha^2}),
\end{align}
which indicates that the third term changes the gauge subsystem state, imparting the QEC capability to the squeezed cat states for photon loss errors. 
Dissipative QEC strategies for squeezed cat codes have been introduced~\cite{xu2023autonomous,shitara2025exploiting} by designing a Lindblad operator $Z_{\rm L} \otimes a_{\rm G}$, which corrects the phase flip in the logical subsystem, followed by the cooling process of the gauge subsystem. 
}

\section{Detail calculation of non-Markovianity for the three-qubit code}
\label{Sec: bitflipp}
We introduce the continuous feedback model described by the GKSL master equation
\begin{equation}
\frac{d}{dt} \rho_{\rm LG}(t) =\mathcal{L}_{\rm LG}[\rho_{\rm LG}(t)]= \gamma  (\sum_m R_m \rho_{\rm LG}(t) R_m ^\dag - \rho_{\rm LG}(t) ),
\end{equation}
where we define $R_m = Q_m^\dag \otimes \ket{0}_G\bra{m}_G$.
Denoting $\mathcal{R}(\cdot)=\sum_m R_m (\cdot) R_m^\dag$, this equation can be easily solved as
\begin{align}
    \rho_{\rm LG}(t) = e^{-\gamma t}\rho_{\rm LG}(0) + (1-e^{-\gamma t})\mathcal{R}[\rho_{\rm LG}(0)].
\end{align}
The steady state in the long-time limit $t\rightarrow \infty$ is identical to the state after the QEC operation $ \rho_{\rm LG}(\infty) =\mathcal{R}[\rho_{\rm LG}(0)] $.
By tracing out the gauge subsystem, we can extract the dynamics of the logical subsystem.
Let the initial state before QEC denote $\rho_{\rm L}(0)=(1-p)\ket{\psi}_{\rm L}\bra{\psi}_{\rm L} + p X \ket{\psi}_{\rm L}\bra{\psi}_{\rm L} X$. 
The steady state denotes $\rho_{\rm L}(\infty)=[1-p^2(3-2p)]\ket{\psi}_{\rm L}\bra{\psi}_{\rm L} + p^2(3-2p) X \ket{\psi}_{\rm L}\bra{\psi}_{\rm L} X$.
Then, tracing out the gauge subsystem yields
\begin{equation}
\begin{aligned}
\rho_{\rm L}(t) &= e^{-\gamma t}  \rho_{\rm L}(0) +(1- e^{-\gamma t}) \rho_{\rm L}(\infty)\\
&=[1-f(t)]\ket{\psi}_{\rm L}\bra{\psi}_{\rm L}+ f(t)X\ket{\psi}_{\rm L}\bra{\psi}_{\rm L}X,
\end{aligned}
\end{equation}
where $f(t)=(1- e^{-\gamma t}) p^2(3-2p)+e^{-\gamma t}p$.
Then, we have
\begin{equation}
\frac{d}{dt} \rho_{\rm L}(t) = -f'(t) (\ket{\psi}_{\rm L}\bra{\psi}_{\rm L} - X \ket{\psi}_{\rm L}\bra{\psi}_{\rm L} X). 
\label{Eq: bit1}
\end{equation}
On the other hand, we have
\begin{equation}
X \rho_{\rm L} (t) X - \rho_{\rm L}(t) = -(1-2f(t)) (\ket{\psi}_{\rm L}\bra{\psi}_{\rm L}-X \ket{\psi}_{\rm L}\bra{\psi}_{\rm L} X ).
\label{Eq: bit2}
\end{equation}
Comparing Eq. \eqref{Eq: bit1} and Eq. \eqref{Eq: bit2}, we obtain
\begin{equation}
\frac{d}{dt}\rho_{\rm L}(t)= \frac{f'(t)}{1-2f(t)} (X \rho_{\rm L}(t) X- \rho_{\rm L}(t)).
\end{equation}
Note that $f'(t) <0$  holds for all $t>0$ when $p<1/2$, because
\begin{equation}
    f'(t)=-\gamma e^{-\gamma t}p(1-p)(1-2p)<0.
\end{equation}
Then, the decay rate measure in Eq. \eqref{Eq: decayrate} for the three-qubit code can be calculated as
\begin{equation}
R(t_1\rightarrow t_3)= \int_{t_1} ^{t_3} \frac{-f'(t)}{1-2f(t)} dt=\frac{1}{2}\log{\left(\frac{1-2f(t_3)}{1-2f(t_1)}\right)}.
\end{equation}

\section{Detail calculation of non-Markovianity for the bosonic code}
We derive the canonical master equation for the bosonic code. The decomposition of the subsystem of the annihilation operator reads
\begin{equation}
V'_{\rm S} a V_{\rm S}^{\prime \dag}= Z_{\rm L} \otimes (\alpha e^{-r} I_{\rm G} + \mathrm{cosh}(r) a_{\rm G}-\mathrm{sinh}(r) a_{\rm G}^\dag )+O(e^{-2\alpha^2}).
\end{equation}
The corresponding state is in the composite Hilbert space consisting of the logical subsystem and gauge subsystem. We denote the general quantum states as $|\psi \rangle=\sum_{ij} c_{ij} |i\rangle_{\rm L} \otimes |j \rangle_{\rm G}$. Now, the annihilation operator operates on the gauge subsystem as $a_{\rm G}|j\rangle_{\rm G}= \sqrt{j} |j-1 \rangle_{\rm G}$ and $a_{\rm G}| 0\rangle_{\rm G}=0$.
Now, the displacement operator can be rewritten with the subsystem decomposition as
\begin{equation}
\begin{aligned}
V'_{\rm S} D(\beta) V_{\rm S}^{\prime \dag}&= V'_{\rm S} e^{\beta a^\dag - \beta^* a} V_{\rm S}^{\prime \dag} \\
&\sim e^{2 i \alpha \mathrm{Im}(\beta) Z_{\rm L}\otimes I_{\rm G}} \times e^{Z_{\rm L} \otimes (\Lambda a_{\rm G}^\dag-\Lambda^* a_{\rm G})},
\end{aligned}
\end{equation}
where $\Lambda= e^r \mathrm{Re}(\beta) +i e^{-r}\mathrm{Im}(\beta)$ with $I_{\rm G}$ being the identity operator of the gauge subsystem. For the initial state represented by the subsystem decomposition $\ket{\psi}_{\rm init}= (c_0 \ket{0}_{\rm L} + c_1 \ket{1}_{\rm L}) \otimes \ket{0}_{\rm G}$, we get
\begin{equation}
\begin{aligned}
\ket{\psi(\Lambda)}:=V_{\rm S}' D(\beta) V_{\rm S}^{\prime \dag} \ket{\psi}_{\rm init} &= c'_0 \ket{0}_{\rm L}\otimes \ket{\Lambda}_{\rm G}+c'_1 \ket{0}_{\rm L}\otimes \ket{-\Lambda}_{\rm G}, \\
\label{Eq: noisesqcat}
\end{aligned}
\end{equation}
and
\begin{align}
\ket{\psi(\Lambda)}\bra{\psi(\Lambda)}=& |c'_0|^2 \ket{0}_{\rm L}\bra{0}_{\rm L} \otimes \ket{\Lambda}_{\rm G}\bra{\Lambda}_{\rm G} 
+ c_0^{ \prime} c_1^{\prime *} \ket{0}_{\rm L}\bra{1}_{\rm L} \otimes \ket{\Lambda}_{\rm G}\bra{-\Lambda}_{\rm G} \notag\\
&+ c_0^{ \prime *} c_1^{\prime } \ket{1}_{\rm L}\bra{0}_{\rm L} \otimes \ket{-\Lambda}_{\rm G}\bra{\Lambda}_{\rm G} + |c'_1|^2 \ket{1}_{\rm L}\bra{1}_{\rm L} \otimes \ket{-\Lambda}_{\rm G}\bra{-\Lambda}_{\rm G},
\label{Eq: rholambda2}
\end{align}
where $e^{2 i \mathrm{Im}(\beta) e^{-r} \alpha' Z_{\rm L} }(c_0 \ket{0}_{\rm L} + c_1 \ket{1}_{\rm L})=c'_0 \ket{0}_{\rm L}+c'_1 \ket{1}_{\rm L}$ and $\ket{\Lambda}_{\rm G}$ is a coherent state for a complex number $\Lambda$. Eq. \eqref{Eq: noisesqcat} indicates that $e^{Z_{\rm L} \otimes (\Lambda a_{\rm G}^\dag -\Lambda^* a_{\rm G})}$ entangles the logical and gauge subsystems. Taking the partial trace for the gauge subsystem for $\ket{\psi(\Lambda)}\bra{\psi(\Lambda)}$ leads to
\begin{equation}
\begin{aligned}
\rho_\Lambda =& |c'_0|^2 \ket{0}_{\rm L}\bra{0}_{\rm L}+c'_0c_1^{\prime *} e^{-2|\Lambda|^2}\ket{0}_{\rm L}\bra{1}_{\rm L} +c_0^{\prime *} c'_1e^{-2|\Lambda|^2}\ket{1}_{\rm L}\bra{0}_{\rm L} + |c'_1|^2 \ket{1}_{\rm L}\bra{1}_{\rm L}.
\label{Eq. rholambda}
\end{aligned}
\end{equation}
Now, we apply the dissipation by $Z_{\rm L} \otimes a_{\rm G}$ in the subsystem decomposition for the autonomous QEC~\cite{xu2023autonomous,shitara2025exploiting}, whose dynamics is described by a \ts{GKSL} master equation
\begin{equation}
\begin{aligned}
\frac{d}{dt} \rho_{  LG}(t) = \frac{\gamma}{2} (2A \rho_{  LG}(t)A^\dag -A^\dag A \rho_{  LG}(t)- \rho_{ LG}(t) A^\dag A),
\end{aligned}
\end{equation}
with $A= Z_{\rm L} \otimes a_{\rm G}$. Then, $\rho_{\rm  LG}(t)$ reads
\begin{equation}
\rho_{\rm LG}(t) = \sum_{k=0}^\infty \frac{(1-e^{-\gamma t})^k}{k!} e^{-\frac{\gamma t}{2} A^\dag A} A^k \rho_{\Lambda} A^{\dag k} e^{-\frac{\gamma t}{2} A^\dag A}.
\end{equation}
We see how the dissipation effect works on each term in $\ket{\psi(\Lambda)}\bra{\psi(\Lambda)}$.
\begin{equation}
\begin{aligned}
&\sum_{k=0}^\infty \frac{(1-e^{-\gamma t})^k}{k!} e^{-\frac{\gamma t}{2} A^\dag A} A^k \ket{0}_{\rm L}\bra{0}_{\rm L} \otimes \ket{\Lambda}_{\rm G}\bra{\Lambda}_{\rm G} A^{\dag k} e^{-\frac{\gamma t}{2} A^\dag A} \\
&=\ket{0}_{\rm L}\bra{0}_{\rm L} \otimes \sum_{k=0}^\infty \frac{(1-e^{-\gamma t})^k}{k!} e^{-\frac{\gamma t}{2} a_{\rm G}^\dag a_{\rm G}} a_{\rm G}^k \ket{\Lambda}_{\rm G}\bra{\Lambda}_{\rm G} a_{\rm G}^{\dag k} e^{-\frac{\gamma t}{2} a_{\rm G}^\dag a_{\rm G}} \\
&= \ket{0}_{\rm L}\bra{0}_{\rm L} \otimes \ket{\Lambda e^{-\gamma t/2}}_{\rm G}\bra{\Lambda e^{-\gamma t/2}}_{\rm G}.
\end{aligned}
\end{equation}
Here, we used $A \ket{0}_{\rm L}\otimes \ket{\phi}_{\rm G} = \ket{0}_{\rm L} \otimes a_{\rm G} \ket{\phi}_{\rm G} $ for a gauge state $\ket{\phi}_{\rm G}$ and
\begin{equation}
\begin{aligned}
&\sum_{k=0}^\infty \frac{(1-e^{-\gamma t})^k}{k!} e^{-\frac{\gamma/2 t}{2} a^\dag a_{\rm G}} a^k \ket{\beta} \bra{\beta} a^{\dag k} e^{-\frac{\gamma/2 t}{2} a^\dag a} = \ket{\beta e^{-\gamma t/2}}\bra{\beta e^{-\gamma t/2}}
\end{aligned}
\end{equation}
for a coherent state $\ket{\beta},~\beta \in \mathbb{C}$. We can perform a similar calculation for $\ket{1}_{\rm L}\bra{1}_{\rm L} \otimes \ket{-\Lambda}_{\rm G}\bra{-\Lambda}_{\rm G}$. Next,
\begin{equation}
\begin{aligned}
&\sum_{k=0}^\infty \frac{(1-e^{-\gamma t})^k}{k!} e^{-\frac{\gamma t}{2} A^\dag A} A^k \ket{0}_{\rm L}\bra{1}_{\rm L} \otimes \ket{\Lambda}_{\rm G}\bra{-\Lambda}_{\rm G} A^{\dag k} e^{-\frac{\gamma t}{2} A^\dag A} \\
&= \ket{0}_{\rm L} \bra{1}_{\rm L} \otimes \sum_{k=0}^\infty \frac{(1-e^{-\gamma t})^k}{k!} (-1)^k (-|\Lambda|^{2k}) e^{-\gamma t/2 a_{\rm G}^\dag a_{\rm G}}  \ket{\Lambda}_{\rm G}\bra{-\Lambda}_{\rm G} a_{\rm G}^{\dag k} e^{-\gamma t/2 a_{\rm G}^\dag a_{\rm G}} \\
&= \bigg[\sum_{k=0}^\infty \frac{( |\Lambda|^2(1-e^{-\gamma t}))^k}{k!}  \bigg]  e^ {-|\Lambda|^2 (1-e^{-\gamma t})} \ket{0}_{\rm L} \bra{1}_{\rm L} \otimes  \ket{\Lambda e^{-\gamma t/2}}_{\rm G}\bra{-\Lambda e^{-\gamma t/2}}_{\rm G} \\
&= \ket{0}_{\rm L} \bra{1}_{\rm L} \otimes  \ket{\Lambda e^{-\gamma t/2}}_{\rm G}\bra{-\Lambda e^{-\gamma t/2}}_{\rm G}.
\end{aligned}
\end{equation}
In the second line, we used $A \ket{0/1}_{\rm L}\otimes \ket{\phi}_{\rm G} = \pm \ket{0/1}_{\rm L} \otimes \ket{\phi}_{\rm G}$. In the third line, we used $e^{-\theta a^\dag a} \ket{\beta} = e^{-|\beta|^2/2 (1-e^{-2\theta})} \ket{\beta e^{-\theta}}$ for $\theta \in \mathbb{R}$. We can perform a similar calculation for $\ket{0}_{\rm L}\bra{0}_{\rm L} \otimes \ket{\Lambda}_{\rm G}\bra{\Lambda}_{\rm G}$. Finally, we obtain
\begin{equation}
\begin{aligned}
\rho_{\rm LG} (t) &= |c'_0|^2 \ket{0}_{\rm L}\bra{0}_{\rm L} \otimes \ket{\Lambda e^{-\gamma t/2}}_{\rm G}\bra{\Lambda e^{-\gamma t/2}}_{\rm G} + c_0^{ \prime} c_1^{\prime *} \ket{0}_{\rm L}\bra{1}_{\rm L} \otimes \ket{\Lambda e^{-\gamma t/2}}_{\rm G}\bra{-\Lambda e^{-\gamma t/2}}_{\rm G} \\
&+ c_0^{ \prime *} c_1^{\prime } \ket{1}_{\rm L}\bra{0}_{\rm L} \otimes \ket{-\Lambda e^{-\gamma t/2}}_{\rm G}\bra{\Lambda e^{-\gamma t/2}}_{\rm G} + |c'_1|^2 \ket{1}_{\rm L}\bra{1}_{\rm L} \otimes \ket{-\Lambda e^{-\gamma t/2}}_{\rm G}\bra{-\Lambda e^{-\gamma t/2}}_{\rm G}.
\label{Eq: rholgtime}
\end{aligned}
\end{equation}
Now, by tracing out the gauge subsystem, we obtain
\begin{equation}
\begin{aligned}
\rho_{\rm L}(t) &= |c'_0|^2 \ket{0}_{\rm L}\bra{0}_{\rm L} + c_0^{ \prime} c_1^{\prime *} e^{-2 |\Lambda|^2 e^{-\gamma t}}\ket{0}_{\rm L}\bra{1}_{\rm L}+ c_0^{ \prime *} c_1^{\prime } e^{-2 |\Lambda|^2 e^{-\gamma t}}\ket{1}_{\rm L}\bra{0}_{\rm L} + |c'_1|^2 \ket{0}_{\rm L}\bra{0}_{\rm L} .
\end{aligned}
\end{equation}
We can easily confirm that $\rho_{\rm L}(t)$ satisfies the following time derivative equation:
\begin{equation}
\frac{d}{dt} \rho_{\rm L}(t)=-|\Lambda|^2 \gamma e^{-\gamma t} (Z_{\rm L} \rho_{\rm L}(t)Z_{\rm L} -\rho_{\rm L}(t) ),
\end{equation}
which have a negative decay rate $-|\Lambda|^2 \gamma e^{-\gamma t}<0$ over the time evolution.

\section{Derivation of canonical master equation for quantum teleportation}

Here, we derive the dynamical equation for the continuous measurement and feedback model of quantum teleportation. We consider the continuous feedback model described by
\begin{equation}
\frac{d}{dt} \rho_{\rm LG}(t) = \gamma  \left[ \left(\sum_m R_m \rho_{\rm LG}(t) R_m ^\dag\right) - \rho_{\rm LG}(t) \right],
\label{Eq: Lindrecov}
\end{equation}
with
\begin{equation}
R_m = Q_{m} ^\dag \otimes \ket{0}_{\rm G}\bra{m}_{\rm G},
\end{equation}
where $\rho_{\rm LG}(t)$ is the state under the subsytem frame and $\gamma$ is a positive constant. For quantum teleportation, we set $m=(m_1, m_2)$ and $R_m=Q_{m_1 m_2}^\dag\otimes \ket{00}\bra{m_1m_2}$.  
We can easily confirm that the following solution satisfies Eq. \eqref{Eq: Lindrecov}: 
\begin{align}
\rho_{\rm LG}(t)&=e^{-\gamma t} \rho_{\rm LG}(0) +(1-e^{-\gamma t})\sum_m R_m \rho_{\rm LG}(0)R_m^\dag\\
&= e^{-\gamma t} \rho_{\rm LG}(0) +(1-e^{-\gamma t}) \ket{\psi}_{\rm L}\bra{\psi}_{\rm L} \otimes \ket{0}_{\rm G}\bra{0}_{\rm G}.
\label{Eq: compositelind}
\end{align}
Here, $\ket{\psi}_{\rm L}$ is the steady state for $t \rightarrow \infty$. Then, tracing out the gauge subsystem yields
\begin{equation}
\rho_{\rm L}(t) = e^{-\gamma t} \rho_{\rm L}(0) +(1-e^{-\gamma t}) \ket{\psi}_{\rm L}\bra{\psi}_{\rm L}.
\label{Eq: logicalsub}
\end{equation}
We can easily confirm that $\rho_{\rm L}(t)$ follows the time derivative equation:
\begin{equation}
\frac{d \rho_{\rm L} (t)}{dt} = \gamma (\ket{\psi}_{\rm L}\bra{\psi}_{\rm L}- \rho_{\rm L}(t)).
\label{Eq: substitute}
\end{equation}
With Eq. \eqref{Eq: logicalsub}, we have for $t>0$
\begin{equation}
\ket{\psi}_{\rm L}\bra{\psi}_{\rm L}= \frac{\rho_{\rm L}(t)-e^{-\gamma t} \rho_{\rm L}(0)}{1- e^{-\gamma t}}.
\label{Eq: target}
\end{equation}
Substituting Eq. \eqref{Eq: target} in Eq. \eqref{Eq: substitute} leads to 
\begin{equation}
\frac{d}{dt} \rho_{\rm L}(t)= \gamma \frac{e^{-\gamma t}}{1-e^{-\gamma t}} (\rho_{\rm L}(t) - \rho(0)).
\label{Eq: timederivative2}
\end{equation}

Then, by regarding Bob's subsystem as a logical subsystem, we have $\rho_{\rm L}(0)=I/2$. Because we have
\begin{equation}
\frac{I}{2} = \frac{\rho_{\rm L}(t)+X_{\rm L} \rho_{\rm L}(t)X_{\rm L} + Y_{\rm L} \rho_{\rm L}(t) Y_{\rm L}+Z_{\rm L} \rho_{\rm L}(t)Z_{\rm L}}{4}
\label{Eq: identity}
\end{equation}
and substituting Eq. \eqref{Eq: identity} in Eq. \eqref{Eq: timederivative2} yields
\begin{equation}
\frac{d \rho_{\rm L}(t)}{dt} = - \frac{\gamma e^{-\gamma t}}{4(1-e^{-\gamma t})} \bigg[\mathcal{L}[X_{\rm L}](\rho_{\rm L}(t))+\mathcal{L}[Y_{\rm L}](\rho_{\rm L}(t))+\mathcal{L}[Z_{\rm L}](\rho_{\rm L}(t)) \bigg],
\end{equation}
where $\mathcal{L}[P](\rho) = P(\rho) P - \rho$ for Pauli operators $P$.

\end{document}